\documentclass[
  journal=largetwo,
  manuscript=article-type,
  year=2020,
  volume=37,
]{cup-journal}

\usepackage{amsmath}
\usepackage{amssymb}
\usepackage{aas_macros}
\usepackage[nopatch]{microtype}
\usepackage{booktabs}

\title{Properties of magnetic turbulence in GRB afterglows}

\author{Evgeny  Derishev}
\affiliation{Institute of Applied Physics, 46 Ulyanov st., 603950 Nizhny Novgorod, Russia}
 %% First letter not capped

\begin{document}

\begin{abstract}

We present a model-independent way to characterise properties of the magnetic-field turbulence in the emitting regions of Gamma-Ray Burst afterglows.
Our only assumption is that afterglows' synchrotron radiation is efficient. It turns out that the gyroradius of plasma particles must be smaller (with a good margin) than the correlation length of the magnetic-field fluctuations. Such turbulence is essentially non-linear and therefore must be produced by some kind of MHD instability, likely acting on top of kinetic Weibel instability. We also find that the emitting particles are loosely confined to local magnetic-field structures and diffusion allows them to sample the entire distribution of local magnetization values. 
This means that one-zone approach to modelling the afterglow spectra is still valid despite the non-linear nature of the magnetic turbulence.
However, the non-linear turbulence may (and likely will) change the synchrotron spectrum of individual electrons.

\end{abstract}

%%%%%%%%%%%%%%%%%%%%%%%%%%%%%%%%%%%%%%%%%%%%%%%
\section{Introduction} 
\label{sec:intro}
%%%%%%%%%%%%%%%%%%%%%%%%%%%%%%%%%%%%%%%%%%%%%%%

Gamma-Ray Burst (GRB) afterglows are emission of relativistic shock waves formed as material ejected by the GRB's central engine hits the circumburst medium. 
This complicated phenomenon could be broadly decomposed into three problems: the emission mechanism, the particle acceleration mechanism, and the shock's microphysics. 

Physics of the emission process itself is generally well understood --- it finds explanation within the framework of the synchrotron-self-Compton (SSC) model \citep[see, e.g.][]{SariPiranNarayan1998,SariEsin2001,NakarAndoSari2009}. The SSC model describes the bulk of afterglow spectrum (up to tens of GeV in photon energy) as synchrotron emission of energetic electrons (injected from some acceleration mechanism) in the magnetic field of the afterglow's shock wave. Some of the synchrotron photons are upscattered by the same electrons to produce another, inverse Compton (IC) component, which extends to TeV energies. The available broad-band observations of GRB afterglows \citep{GRB190114C,GRB190829A} show that the synchrotron component dominates over the IC component. Hence, we do not include the latter in our analysis.

There are two options for the particle acceleration mechanism in relativistic shocks: a modification of common diffusive shock acceleration (DSA) \citep[see, e.g.][]{Achterberg2001,KeshetWaxman2005} and the converter mechanism \citep{Converter}, whose specific realisation for the SSC-emitting shocks is the pair-balance model \citep{PairBalance}. In the pair-balance model the accelerated electrons (and positrons) are $e^{-}e^{+}$-pairs, produced in the upstream via collisions of IC photons with synchrotron photons, and then transported by the flow into the downstream gaining energy thereby. 
The pair-balance model makes several predictions that set it apart from models based on DSA.
Most notably, the energy of accelerated leptons is predicted to increase as the shock decelerates, the ratio of IC to synchrotron power is of the order of unity for all the afterglows regardless of their parameters and at all temporal phases, and position of the IC peak remains approximately constant over time (for wind-like environment). All of these a priori claims were confirmed few years later, when TeV observation became available.

The shock's microphysics, and especially the process of magnetic field generation in initially weakly magnetized upstream fluid, is the least clear part of the puzzle. Yet it has implications for both the emission mechanism and the particle acceleration mechanism. Numeric simulations of collisionless and initially unmagnetized relativistic shocks successfully demonstrated generation of the magnetic field turbulence at the shock front, but this magnetic field is very short-lived compared to the synchrotron-loss timescale \citep[see, e.g.][]{ChangSpitkovskyArons2008,SironiKeshetLemoine2015}. The pair-balance model may help resolving this issue --- its distinctive feature is greatly extended in time injection of anisotropic particles into the upstream, that slows growth of the magnetic turbulence and potentially makes it persistent over the entire synchrotron cooling time. The possibility that slower growth of the magnetic turbulence leads to longer lifetime, suggested in the original paper, was later confirmed numerically \citep{SlowDecay}. However, it remains unclear if a kinetic instability with the growth rate that matches the actual afterglow durations is capable of producing strong enough magnetic turbulence.

In this paper we present a way to characterize the magnetic turbulence in GRB afterglows using most general and model-independent approach. 
To keep our analysis more general, we treat the post-shock fluid as consisting of two components: the synchrotron-emitting electrons that produce the observed radiation, and the plasma particles (either ions or the emitting electrons themselves) that takes up most of the energy density and conveys currents supporting the magnetic field.
Such a system is characterized by 6 parameters: plasma frequency $\omega_\mathrm{pl}$ (or, equivalently, plasma skin depth $\ell_s = c/\omega_\mathrm{pl}$), correlation length of the magnetic field fluctuations $\lambda_c$, decay time of the turbulent magnetic field $t_\mathrm{dec}$, gyroradius of plasma particles $r_\mathrm{g,pl}$, radiative energy loss timescale for emitting electrons $t_\mathrm{cool}$, and their gyroradius $r_\mathrm{g,e}$. There are 5 independent equations for these quantities; they are given in Sec.~(\ref{sec:main}).
Thus, there exist pairwise relations between the aforementioned parameters. 

The most interesting is the relation between the gyroradius of plasma particles and typical scale of the magnetic field fluctuations. If $r_\mathrm{g,pl}$ turns out to be larger than $\lambda_c$, then the current-bearing particles freely move between fluctuations and the magnetic turbulence is linear. Otherwise, in the case $r_\mathrm{g,pl} < \lambda_c$, most of the plasma particles are trapped in regions of stronger magnetic field and the turbulence is non-linear. Linear turbulence can result from any instability (or combination of instabilities), both kinetic and MHD. Conversely, non-linear turbulence ultimately points at some MHD mechanism, whereas kinetic instabilities can play only auxiliary role. 

Surprisingly, under a single assumption that synchrotron radiation is efficient in GRB afterglows, we are able to prove a rather strong statement that the magnetic turbulence in afterglow emitting regions is non-linear. The logic of this paper is proof by contradiction. Starting from dispersion relation for linear magnetic turbulence, we arrive at estimate of $r_\mathrm{g,pl} / \lambda_c$ ratio whose value points at strong non-linearity, thus proving that the magnetic turbulence in GRB afterglow regions is essentially non-linear. Same claim can be formulated in other words: by requiring a linear turbulence, one effectively places an upper limit on the synchrotron efficiency, which turns out to be far too low to account for afterglow observations.

Our finding is unexpected and likely means that the present-day theories of shocks' microphysics need a significant modification.
This has serious implications for understanding physics of relativistic shocks in general, and for GRB afterglow theories in particular.

%%%%%%%%%%%%%%%%%%%%%%%%%%%%%%%%%%%%%%%%%%%%%%%
\section{Relations between shock's microscopic parameters} 
\label{sec:main}
%%%%%%%%%%%%%%%%%%%%%%%%%%%%%%%%%%%%%%%%%%%%%%%

Efficient synchrotron emission implies that the magnetic field does not decay faster than electrons lose their energy radiatively. So, the source must satisfy 
\begin{equation} \label{EfficiencyCondition}
    t_\mathrm{dec} \gtrsim t_\mathrm{cool} \ .
\end{equation}

Decay time for the turbulent magnetic field depends on the typical wavenumber of the fluctuations (i.e. on their inverse correlation length $\lambda_c$) and can be estimated from dispersion relation for Weibel instability modes assuming zero anisotropy
\citep{MorseNielson1971}\footnote{We take Eq.~(6) from this paper in the limit, where $\omega/(k_y v_y) \ll 1$, thermal velocity is $v_y = c$, and ignore numerical factor $\pi^{-1/2}$. Note that $k_y$ in their notation is equivalent to our $\lambda_c^{-1}$. Since the solution of their dispersion relation is aperiodic, 
their $\omega$ is the same as $t_\mathrm{dec}^{-1}$.}: 
\begin{equation} \label{TurbulenceDecayTime}
    t_\mathrm{dec} 
    \simeq  \frac{\omega_\mathrm{pl}^2 \lambda_c^3 }{c^3}  \ .
\end{equation}
This relation may break for non-linear turbulence, and it is not clear beforehand whether the decay of non-linear fluctuations will be faster or slower.

The cooling time for synchrotron-radiating electrons is
\begin{equation} \label{CoolingTime}
    t_\mathrm{cool} \simeq \frac{6 \pi \ m_e c}{\gamma \sigma_\mathrm{_T} B^2} 
    = \frac{9}{4} \frac{m_e c^3}{\gamma e^2 \omega_\mathrm{_B}^2} \ ,
\end{equation}
where $B$ is the magnetic field strength, $\gamma$ the Lorenz factor of emitting electrons, $\sigma_\mathrm{_T}$ Thomson cross-section, $m_e$ electron mass, $e$ elementary charge, and $c$ speed of light. The cyclotron frequency
\begin{equation} \label{OmegaB}
    \omega_\mathrm{_B} \simeq \frac{e B}{m_e c} 
\end{equation}
in turn can be expressed in terms of relativistic plasma frequency
\begin{equation} \label{PlasmaFreq}
    \omega_\mathrm{pl} \simeq \left( \frac{4\pi e^2 N c^2}{E_\mathrm{pl}} \right)^{1/2} ,
\end{equation}
where $N$ is number density of plasma particles and $E_\mathrm{pl}$ their energy. The ratio of these frequencies is
\begin{equation} \label{OmegaB-to-PlasmaFreq}
    \frac{ \omega_\mathrm{_B} }{ \omega_\mathrm{pl} } 
    \simeq \frac{E_\mathrm{pl}}{m_e c^2} \left( \frac{B^2 /  4\pi }{ N E_\mathrm{pl} }  \right)^{1/2}
    = \sqrt{2\epsilon_\mathrm{_B}} \ \frac{E_\mathrm{pl}}{m_e c^2} \ ,
\end{equation}
where $\epsilon_\mathrm{_B}$ is the ratio of the magnetic-field energy density to plasma energy density.

Both emitting electrons and plasma particles are relativistic, so that their giroradii are given by similar expressions:
\begin{equation} \label{Gyroradii}
    r_\mathrm{g,e} = \gamma \ \frac{c}{\omega_\mathrm{_B}} \ , \quad
    r_\mathrm{g,pl} = \frac{E_\mathrm{pl}}{m_e c^2} \ \frac{c}{\omega_\mathrm{_B}} \ ,
\end{equation}
and their ratio is
\begin{equation} \label{GyroradiiRatio}
    \frac{r_\mathrm{g,pl}}{r_\mathrm{g,e}} 
    = \frac{E_\mathrm{pl}}{\gamma m_e c^2}  \ .
\end{equation}
The ratio $(\gamma m_e c^2)/E_\mathrm{pl}$ is closely related to the standard equipartition parameter $\epsilon_\mathrm{e}$ (the ratio of total energy of radiating particles to the total downstream energy). 
One normally expects $(\gamma m_e c^2)/E_\mathrm{pl} \gtrsim \epsilon_\mathrm{e}$ because the number of accelerated leptons is usually less than the number of protons. This is true for both DSA (not all the ambient electrons are accelerated, and the flow just drags some of them into the downstream) and pair balance model (though in the very early afterglow the number of secondary pairs could be comparable to the number of ambient electrons).

Equations (\ref{EfficiencyCondition}), (\ref{TurbulenceDecayTime}), (\ref{CoolingTime}), (\ref{OmegaB-to-PlasmaFreq}), and (\ref{GyroradiiRatio}) are the 5 independent equations relating the 6 unknown quantities of the problem.
In addition, it is convenient to express the product $(\omega_\mathrm{_B}/\gamma) t_\mathrm{cool}$ using observable quantities:
\begin{equation} \label{Gyrofreq-times_CoolingTime}
    (\omega_\mathrm{_B}/\gamma) t_\mathrm{cool} \simeq 
    \frac{9}{4} \frac{m_e c^3}{\gamma^2 e^2 \omega_\mathrm{_B}}  =
    \frac{9}{4} \frac{\Gamma m_e c^2}{\alpha_f E_\mathrm{sy}} \ ,
\end{equation}
where $\alpha_f$ is the fine-structure constant and 
\begin{equation} \label{ObservedPeakEnergy}
    E_\mathrm{sy} \simeq \Gamma \gamma^2 \hbar \omega_\mathrm{_B}
\end{equation}
the observed energy of photons in the synchrotron peak. 
We also consider the shock's $\Gamma$ to be an observable, as it allows determination from spectrum and lightcurve modelling in a fairly independent way.

Now we re-write the efficiency condition (\ref{EfficiencyCondition}) in terms of $(r_\mathrm{g,pl}/\lambda_c)$ ratio using Eq.~(\ref{TurbulenceDecayTime}) to replace $t_\mathrm{dec}$:
\begin{equation} 
    \frac{r_\mathrm{g,pl}^3}{ \lambda_c^3} \lesssim \frac{\omega_\mathrm{pl}^2 \ r_\mathrm{g,pl}^3}{c^3 \ t_\mathrm{cool}} 
    = \frac{\gamma^3 \omega_\mathrm{pl}^2}{\omega_\mathrm{_B}^3  t_\mathrm{cool}} \left( \frac{r_\mathrm{g,pl}}{r_\mathrm{g,e}} \right)^3 \ .
\end{equation}
Substituting $(\omega_\mathrm{_B}/\omega_\mathrm{pl})$ from Eq.~(\ref{OmegaB-to-PlasmaFreq}), $(r_\mathrm{g,pl}/r_\mathrm{g,e})$ from Eq.~(\ref{GyroradiiRatio}), and $(\omega_\mathrm{_B}/\gamma) t_\mathrm{cool}$ from Eq.~(\ref{Gyrofreq-times_CoolingTime}) we arrive at
\begin{equation} \label{LinearityEstimate}
    \frac{r_\mathrm{g,pl}}{\lambda_c} 
    \lesssim  \left( \frac{2}{9} \frac{1}{\epsilon_\mathrm{_B}}   \frac{\alpha_f E_\mathrm{sy}}{\Gamma m_e c^2} \right)^{1/3} \left( \frac{E_\mathrm{pl}}{\gamma m_e c^2} \right)^{1/3} .
\end{equation}
We conclude that plasma particles are confined by the magnetic field fluctuations and the magnetic turbulence in GRB afterglow emitting regions is essentially non-linear. Indeed, GRB afterglows typically have peak of their synchrotron emission in the few keV range (up to tens of keV in early afterglows), the only available determination of the magnetization parameter from the complete spectral analysis is 
$\epsilon_\mathrm{_B} \simeq 0.004$ \citep{AsanoMuraseToma2020, GRBparameters}, and the numerical simulations tend to produce  $\epsilon_\mathrm{_B} \sim 10^{-3} \div 10^{-2}$ \citep[e.g.][]{ChangSpitkovskyArons2008,SironiKeshetLemoine2015}. Altogether, the first factor in the above inequality evaluates to $\sim 0.05$. The second factor cannot revert conclusion about smallness of $(r_\mathrm{g,pl}/\lambda_c)$ ratio --- this would require unreasonably low energy of emitting electrons
(implying extremely low value of $\epsilon_\mathrm{e}$),
at least four orders of magnitude lower than the energy of plasma particles (a standard expectation is $(\gamma m_e c^2)/E_\mathrm{pl} \gtrsim 1$).
We remind that Eq.~(\ref{LinearityEstimate}) is derived under assumption of high radiative efficiency and hence may not be valid for very late afterglow, when the radiative efficiency likely experiences significant decline.

Another, more general, way of expressing our conclusions is to evaluate the upper limit for synchrotron radiative efficiency for a linear magnetic turbulence, i.e., under condition $r_\mathrm{g,pl} > \lambda_c$. For a low synchrotron efficiency, $f_\mathrm{sy} \ll 1$, Eq.~(\ref{EfficiencyCondition}) takes the form  $t_\mathrm{dec} = f_\mathrm{sy} t_\mathrm{cool}$. Propagating the factor $f_\mathrm{sy}$ through all the equations, we obtain from Eq.~(\ref{LinearityEstimate})
\begin{equation} \label{EfficiencyLimit}
    f_\mathrm{sy} \lesssim 
    \frac{2}{9} \frac{1}{\epsilon_\mathrm{_B}}   \frac{\alpha_f E_\mathrm{sy}}{\Gamma m_e c^2}  
    \left( \frac{E_\mathrm{pl}}{\gamma m_e c^2} \right) \sim 10^{-4}  \left( \frac{E_\mathrm{pl}}{\gamma m_e c^2} \right) \ ,
\end{equation}
where we take $\lambda_c/r_\mathrm{g,pl}=1$ as the upper limit for this ratio for a linear turbulence. The limit set by Eq.~(\ref{EfficiencyLimit}) is far too low to account for the majority of GRB afterglows, thus once again stressing the necessity to have non-linear magnetic turbulence in afterglows' blast waves.

%%%%%%%%%%%%%%%%%%%%%%%%%%%%%%%%%%%%%%%%%%%%%%%
\section{Implications} 
\label{sec:implications}
%%%%%%%%%%%%%%%%%%%%%%%%%%%%%%%%%%%%%%%%%%%%%%%

There is common agreement that relativistic shocks propagating into unmagnetized medium (as in the case of GRB afterglows) generate their own magnetic field through Weibel-type (filamentation) instability, as suggested by \cite{MedvedevLoeb1999}. The actual properties of the magnetic turbulence, such as the spatial scale of fluctuations and the level of magnetization, depend on what is the source of driving anisotropy in particles' distribution. For example, predictions of the common collisionless shock scenario largely differ from predictions of the pair balance model, as discussed in the introduction. More important, however, is the fact that Weibel-type instability always saturates before entering non-linear phase \citep{LinearWeibel} and therefore cannot explain non-linear turbulence required for GRB afterglows. This means that our understanding of processes that lead to generation of the magnetic field in relativistic shocks is essentially incomplete. Whatever is the outcome of Weibel-type instability, it should be followed by some kind of MHD instability capable of reaching the required level of non-linearity. 
And it is the latter that ultimately determines the level of magnetization. It is too early to speculate about the nature of this MHD instability.

In the pair-balance model, the magnetic field builds up due to continuous injection of electron-positron pairs, which occurs on the timescale approximately equal to $t_\mathrm{cool}$. A natural expectation is that the magnetic-field decay time is comparable to the build-up time, suggesting order-of-magnitude equality $t_\mathrm{cool} \sim t_\mathrm{dec}$ instead of relation (\ref{EfficiencyCondition}). If so, then Eq.~(\ref{LinearityEstimate}) should be treated as an estimate for the $(r_\mathrm{g,pl}/\lambda_c)$ ratio rather than the upper limit for it.

Generally speaking, one should not treat our result as a proof that the magnetic-field turbulence is strongly non-linear. It is possible that the actual fluctuations are just non-linear enough to form soliton-like structures, which could persist much longer than linear perturbations of the same scale.

Another important question is whether the magnetic turbulence is linear with respect to the emitting electrons. As follows directly from Eq.~(\ref{LinearityEstimate}),
via substitution of $(r_\mathrm{g,pl}/r_\mathrm{g,e})$ from Eq.~(\ref{GyroradiiRatio}),
\begin{equation} \label{LinearityEstimate2}
    \frac{r_\mathrm{g,e}}{\lambda_c}
    \lesssim  \left( \frac{2}{9} \frac{1}{\epsilon_\mathrm{_B}}   \frac{\alpha_f E_\mathrm{sy}}{\Gamma m_e c^2} \right)^{1/3} \left( \frac{\gamma m_e c^2}{E_\mathrm{pl}} \right)^{2/3} .
\end{equation}
This indicates that for the emitting electrons the turbulence is still non-linear at least at the early afterglow stage.
The pair-balance model predicts increase of $\gamma$ with time, roughly following $B^{-1/3}$ law, while $E_\mathrm{pl}$ is expected to decrease proportionally to the shock's Lorentz factor $\Gamma$. Then, for the late stages of afterglow evolution, one may expect that the turbulence becomes linear with respect to the emitting electrons (being non-linear with respect to the plasma particles).

In any case, the emitting electrons are loosely confined in the magnetic field inhomogeneities. Electrons' displacement due to diffusion can be estimated as
\begin{equation} \label{DiffusionDisplacement}
    \ell_\mathrm{diff} \simeq \sqrt{D t_\mathrm{cool}} 
    \simeq r_\mathrm{g,e} \sqrt{\omega_\mathrm{_B} t_\mathrm{cool} / (3 \gamma)} \ ,
\end{equation}
where we assumed Bohm diffusion coefficient $D=c r_\mathrm{g,e}/3$. Substituting $r_\mathrm{g,e}$ from Eq.~(\ref{LinearityEstimate2}) and $(\omega_\mathrm{_B}/\gamma) t_\mathrm{cool}$ from Eq.~(\ref{Gyrofreq-times_CoolingTime}), we find that
\begin{equation} \label{DiffusionDisplacement2}
    \frac{\ell_\mathrm{diff}}{\lambda_c}  
    \lesssim \left( \frac{1}{48} \frac{1}{\epsilon_\mathrm{_B}^2} \frac{\Gamma m_e c^2}{\alpha_f E_\mathrm{sy}}  \right)^{1/6}
    \left( \frac{\gamma m_e c^2}{E_\mathrm{pl}} \right)^{2/3} 
    \sim 30 \left( \frac{\gamma m_e c^2}{E_\mathrm{pl}} \right)^{2/3} .
\end{equation}
Here we take $\epsilon_\mathrm{_B} = 0.004$, $E_\mathrm{sy} = 5$~keV, and $\Gamma=30$ to estimate the first factor. Thus, the emitting electrons travel diffusively far enough to sample many irregularities of the magnetic field over their cooling time. This means that one-zone approximation for GRB afterglow emitting region can still be used despite non-linear character of the magnetic turbulence. However, it is necessary to average the synchrotron spectrum of an individual electron over regions with different magnetic field strength. It is hard to guess a priori what exactly could be the distribution function for local values of the magnetic field strengths, and this distribution may strongly deviate from Gaussian, expected for a linear turbulence.

Let us approximate fluctuations as spherical structures with a stronger magnetized core of size $R_c$ surrounded by a weaker magnetic field that declines as power-law of distance until joining with a neighbouring fluctuation at $R \simeq R_\mathrm{rim}$. Then the distribution of the magnetic field strength is
\begin{equation} \label{FluctuationStructure}
    B \simeq
    \begin{cases}
      B_c , \quad R<R_c \\
      B_c \left( \frac{R_c}{R} \right)^{k} , \quad R_c < R < R_\mathrm{rim}
    \end{cases} \ .
\end{equation}
An electron, that samples distribution (\ref{FluctuationStructure}) spending equal time in each volume element, produces space-averaged spectral energy distribution (SED, $\nu F_{\nu}$) consisting of two power-law segments followed by an exponential cut-off. 
At frequencies below $\nu_\mathrm{rim} = \gamma_e^2 e B_\mathrm{rim} /\left( 2\pi m_e c \right)$, the spectral index is the same as for an electron in uniform magnetic field, i.e. $\nu F_{\nu} \propto \nu^{4/3}$. The SED of the higher-frequency power-law segment forms in the region of declining magnetic field as product of local synchrotron power (proportional to $B^2$) and the available volume ($\propto R(B)^3$), so that $\nu F_{\nu} \propto \nu^{2-3/k}$. We summarize the spectral shape of synchrotron radiation from an individual electron as
\begin{equation} \label{IndividualSED}
    \nu F_{\nu} \propto
    \begin{cases}
      \nu^{4/3} , \quad \nu < \nu_\mathrm{rim} = \gamma_e^2 \frac{e B_\mathrm{rim}}{ 2\pi m_e c} \\
      \nu^{2-3/k} , \quad \nu_\mathrm{rim} < \nu < \nu_c = \gamma_e^2 \frac{e B_c}{ 2\pi m_e c}  
    \end{cases} \ .
\end{equation}
If $k>4.5$, then the spectral index in the second segment is $4/3$.
The spectrum (\ref{IndividualSED}) cuts off exponentially at $\nu_c$.
 
There are two different cases depending on the value of the decline index $k$.

One case is $k<3/2$. Then the r.m.s. magnetic field strength is $\simeq B_\mathrm{rim}$ and the SED peak of an individual electron is located at frequency 
$\nu_\mathrm{rim}$, which in this case corresponds to typical synchrotron frequency of an electron radiating in r.m.s. magnetic field. At frequencies above the peak the SED extends as a power-law before exponential cut-off at $\nu_c > \nu_\mathrm{rim} \equiv \nu_\mathrm{peak}$. 

The more likely and noteworthy case is $k>3/2$ (the value $k=3$, corresponding to magnetic dipole, seems to be the most realistic). 
In this case the SED peaks at frequency $\nu_c$ and cuts off right after the peak. Before the peak the SED is softer than that of an electron in a uniform magnetic field, provided $k<4.5$. This would rise estimated levels of optical emission in GRB afterglows. Another, potentially even more important circumstance, is shift of SED's peak to a higher frequency compared to the value which corresponds to the r.m.s. magnetic field. Indeed, the r.m.s. magnetic field strength is $B_\mathrm{rms} \simeq B_c \left( R_c / R_\mathrm{rim} \right)^{3/2}$, much less than the core's field strength that determines location of the SED's peak. 
One may expect that in non-linear MHD turbulence plasma beta drops to a value of the order of unity inside the core's radius $R_c$, then the SED's peak frequency may go up by a factor $\sim \epsilon_\mathrm{_B}^{-1/2}$, i.e. may increase by an order of magnitude or even more.

Interestingly, there are two physically distinct situations previously considered in the literature as theoretical possibilities, which lead to SEDs formally resembling the above two cases. The first case has similarity to the situation where there is magnetic turbulence with power-law distribution of local field strengths \citep{UvarovBykov2023}.
The second case has partial similarity to the situation \citep[see, e.g.][]{KelnerAharonianKhangulyan2013} where the magnetic fluctuations have very small spatial scale, much less than the gyroradius of sub-relativistic electrons.

Our findings have some implications for the particle acceleration process and seem to favor the pair balance model. Indeed, the converter acceleration mechanism (which is the basis for the pair balance model) is insensitive to the fact that leptons are trapped by the magnetic field fluctuations because transport from downstream to upstream occurs via neutrals. On the contrary, non-linear turbulence, in general, prohibits DSA. The only way to keep it a viable possibility is a situation where there is a special hierarchy of spatial scales --- DSA takes place in a normal way on the shortest scale, the non-linear MHD fluctuations develop on a longer scale, and radiation takes place on the longest scale.

%%%%%%%%%%%%%%%%%%%%%%%%%%%%%%%%%%%%%%%%%%%%%%%
\section{Summary} 
\label{sec:summary}
%%%%%%%%%%%%%%%%%%%%%%%%%%%%%%%%%%%%%%%%%%%%%%%

In this paper we show that the magnetic turbulence in GRB afterglow emitting regions must be strongly non-linear (in the sense that gyroradii of current-bearing particles are much smaller than the size of magnetic-field fluctuations) to ensure high synchrotron efficiency. For a linear turbulence, the synchrotron efficiency is limited to $f_\mathrm{sy} \lesssim 10^{-4}$, being far too low to explain observations.
Amplification of the magnetic field by an MHD instability seems to be inevitable requirement for a strongly non-linear turbulence to appear. This MHD instability must develop on top of kinetic Weibel (filamentation) instability, which is known to operate in relativistic collisionless shocks.

The emitting electrons are not confined within the magnetic-field fluctuations and can diffusively travel between them, sampling the entire magnetic field distribution and ensuring validity of the one-zone approximation. However, the space-averaged synchrotron SED of individual electrons is largely distorted compared to SED in a uniform magnetic field. Namely, the SED's peak frequency may shift up by a large factor (an order of magnitude or even more) and the low-frequency asymptotic may be considerably softer. This will significantly alter estimations of the emitting zone parameters and will produce much more optical emission than was expected previously.

%%%%%%%%%%%%%%%%%%%%%%%%%%%%%%%%%%%%%%%%%%%%%%%
\begin{acknowledgement}
% Insert the Acknowledgment text here.
\end{acknowledgement}

\paragraph{Funding Statement}

This work was supported by the Russian Science Foundation under grant no.  24-12-00457.

\paragraph{Competing Interests}

None.

\paragraph{Data Availability Statement}

Data sharing is not applicable to this article as no new data were created or analysed in this study.

%\endnote in some journals will behave like \footnote; and \printendnotes will not output anything. 
\printendnotes

\printbibliography

\end{document}